\documentclass[prl,aps,twocolumn,showpacs,superscriptaddress,nofootinbib]{revtex4}
\usepackage{graphicx}  \usepackage{dcolumn}  \usepackage{bm} \usepackage{amsmath}  \usepackage{amsfonts}  \usepackage{epsfig} \usepackage{fancyhdr}  \usepackage{enumerate}  \usepackage{float} \usepackage{setspace}  \usepackage{fancybox}

\begin{document}

\title{Temporal Quantum Control with Graphene}
\author{Alejandro Manjavacas}
\email{a.manjavacas@csic.es}
\affiliation{IQFR - CSIC, Serrano 119, 28006 Madrid, Spain}
\author{Sukosin Thongrattanasiri}
\affiliation{IQFR - CSIC, Serrano 119, 28006 Madrid, Spain}
\author{Darrick E. Chang}
\affiliation{ICFO-Institut de Ciencies Fotoniques, Mediterranean Technology Park, 08860 Castelldefels (Barcelona), Spain}
\author{F. Javier Garc\'{\i}a de Abajo}
\email{J.G.deAbajo@csic.es}
\affiliation{IQFR - CSIC, Serrano 119, 28006 Madrid, Spain}

\date{\today}

\begin{abstract}
We introduce a novel strategy for controlling the temporal evolution of a quantum system at the nanoscale. Our method relies on the use of graphene plasmons, which can be electrically tuned in frequency by external gates. Quantum emitters (e.g., quantum dots) placed in the vicinity of a graphene nanostructure are subject to the strong interaction with the plasmons of this material, thus undergoing time variations in their mutual interaction and quantum evolution that are dictated by the externally applied gating voltages. This scheme opens a new path towards the realization of quantum-optics devices in the robust solid-state environment of graphene.
\end{abstract}
\maketitle



Controlling the temporal quantum evolution of a physical system by means of external macroscopic stimuli will make versatile quantum-information devices viable \cite{CZK97}. Various methods for quantum control have been proposed \cite{CZK97}, and despite efforts and progress in fields such as ion traps \cite{AUW10,RNH12}, scalable systems remain an open challenge. Recently, solid-state quantum devices are attracting growing interest because they provide a robust platform for implementing scalable temporal control. Here we show that doped graphene nanostructures combined with two-level atoms or quantum dots provide a robust platform for achieving the desired goal of full temporal quantum control. The interaction between the quantum dots is strongly mediated by plasmons in the graphene, which can be electrostatically tuned through engineered gates \cite{LHJ08,CPB11}. The quantum evolution of the dots is then manipulated by modulating over time the electric potential that we apply to the gates. We provide realistic simulations demonstrating excellent control over the decay of individual and interacting dots. Any desired decay profile can be produced by resorting on the unprecedentedly fast electro-optical modulation of graphene \cite{graphene1st} and its strong interaction with neighboring quantum emitters \cite{paper167}. This constitutes a radically new path towards controlling quantum systems in nanoscale solid-state environments by means of conventional electronics, with the capacity of bringing quantum devices closer to reality.

Quantum mechanics rules the temporal evolution at small length and energy scales, giving rise to non-intuitive properties such as quantum superposition and entanglement. These phenomena provide an extra handle to process information \cite{ABL08,K08}, to improve optical metrology and break the imaging diffraction limit \cite{D08}, and to substantially alter the statistics of light \cite{M1986,OTL06,AAS10}, with a vast range of potential applications, including quantum computing and cryptography \cite{NC04} and sensing \cite{D08}. However, controlling the quantum evolution at such length and energy scales remains a challenge that can only be partially addressed by using elaborate setups, for example in the context of cavity-QED \cite{CZK97,ECZ97,RNH12}. Recently, surface plasmons have been identified as a potential candidate to mediate the interaction between externally controlled signals (e.g., laser beams) and small quantum systems (e.g., quantum dots), allowing one to explore exciting phenomena such as single photon transistors \cite{CSD07}, entanglement \cite{GMM11} and quantum blockade \cite{paper184} among other feats. In the following, we show that graphene can also be used to control the evolution of a quantum system interacting with it through a classical electric-potential signal. Specifically, we show that direct control over two-level emitters is possible, thus eliminating the need to involve additional excited or metastable states and external control fields upon which other methods are relying \cite{CZK97,DM08}. Our results can be used for a robust, solid-state implementation of photon storage \cite{GAF07} and cascaded quantum systems in which one system drives the evolution of another system \cite{G93,C93}. These are basic elements needed for distributed quantum networks and processors, and quantum state transfer.

The emergence of graphene as a tunable plasmonic material, in which plasmons can be literally switched on and off by applying external potentials \cite{graphene1st}, opens a natural way to control the quantum evolution of small systems through plasmon-mediated interactions, which are in turn modulated by external fields. Charge carriers in graphene, the so-called massless Dirac Fermions, follow a linear dispersion relation that lead to several exciting properties. Among them, low-energy plasmons exist in the atomically thin carbon film when it is electrically charged, and their frequencies scale as $n^{1/4}$ with the doping electron density \cite{JBS09}  $n$. Therefore, the plasmon frequencies can be controlled by changing $n$, which is in turn proportional to the magnitude of the perpendicular electric field. This field can be supplied by biasing the graphene with respect to a nearby gate, and thus, the voltage applied through the gates directly modifies in a predictable way the plasmon frequencies.

\begin{figure*}
\includegraphics[width=170mm,angle=0]{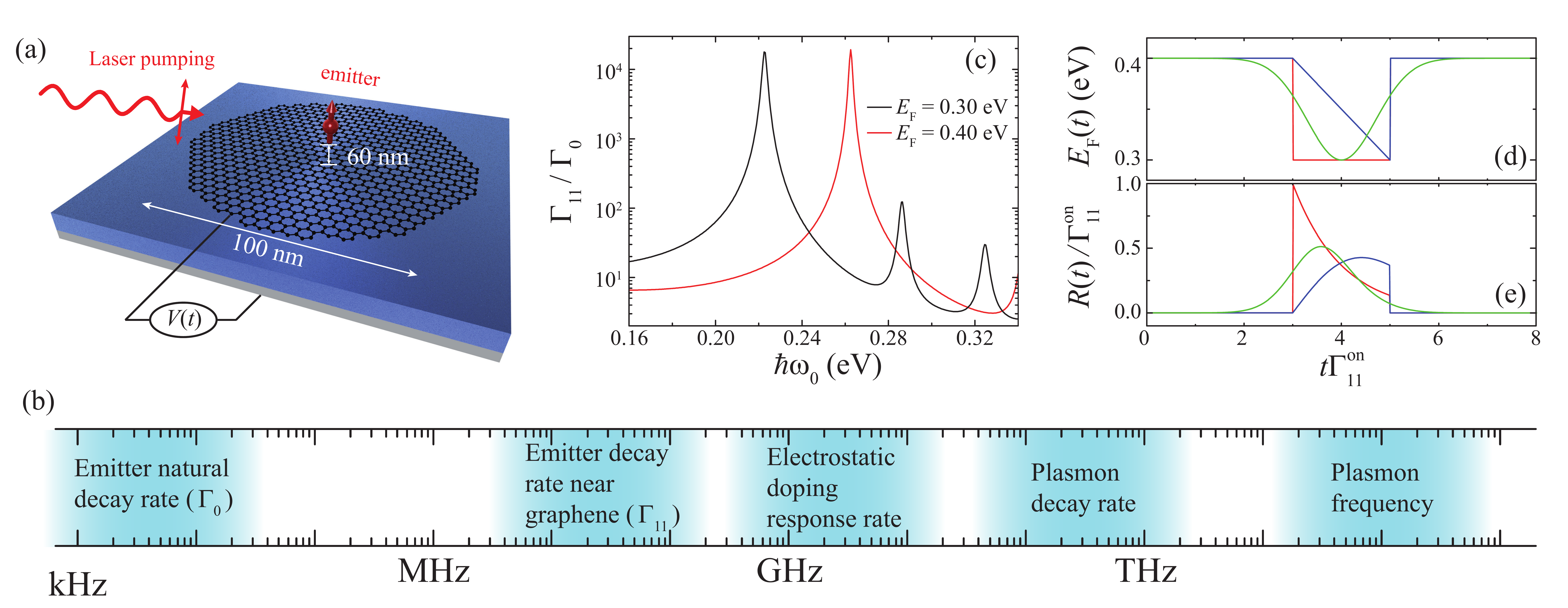}
\caption{Temporal control over the quantum evolution of an optical emitter via interaction with a doped graphene nanostructure. (a) We consider a two-level optical emitter placed above a graphene nanodisk. The emitter is excited by a laser pulse. Electrical doping of the graphene through a bias potential $V$ allows the nanodisk to support plasmons. The plasmon frequency is proportional to $\sqrt{|V|}$, and thus, it can be controlled over time by modulating $V(t)$. The coupling of the emitter to the plasmon and therefore the quantum evolution of the emitter state are both controlled by $V$. (b) Different time scales are involved in the evolution of the emitter-graphene system, represented here through the emitter natural decay rate $\Gamma_0$, the enhanced decay rate $\Gamma_{11}$ produced by interaction with the graphene plasmons, and the plasmon decay rate and frequency typical of doped graphene. It is important to stress that the plasmon lifetime is short compared to the electric modulation rates of the doping potential, in the GHz range, which is in turn fast compared to the lifetime of the excited emitter. (c) The increase in emission rate $\Gamma_{11}/\Gamma_0$ is shown as a function of emission frequency for two different doping levels, quantified through the graphene Fermi energy $E_{\rm F}$. The temporal profile of the emission rate $R(t)$ can be controlled by suitably modulating $E_{\rm F}$ over time as shown in (d) and (e) for $\hbar\omega_0=0.22\,$eV.}
\label{Fig1}
\end{figure*}

In this Letter, we show that the interaction between one or more quantum emitters with graphene plasmons strongly modifies the evolution of the system, which can be temporally controlled by switching on and off the plasmons through electrostatic gating. We first illustrate this concept by analyzing the interaction of a doped graphene nanodisk with a single quantum emitter (Fig.\ \ref{Fig1}). We choose a nanodisk of $100$\,nm in diameter, with a level of doping characterized by a Fermi energy \cite{CGP09} $E_{\rm F}=\hbar v_F\sqrt{\pi |n|}$, where $v_F\approx10^6$m/s is the Fermi velocity. This determines the plasmon energy $\hbar\omega_p\propto\sqrt{E_{\rm F}}$ and the plasmon decay rate \cite{paper176} $\Gamma_p\propto1/E_{\rm F}$. The emitter is placed $60$\,nm above the center of the graphene nanodisk, and is modeled as a two-level system with a characteristic transition energy $\hbar\omega_0$ and a natural decay rate $\Gamma_0$. The different time scales that characterize the evolution of the system are plotted in Fig.\ \ref{Fig1}(b). We choose a value of $\Gamma_0\sim10^3-10^4$\,s$^{-1}$ typical of slowly emitting atoms such as erbium. When the emitter is placed close to the nanodisk, its decay rate is enhanced up to $\Gamma_{11}\sim10^4\,\Gamma_0$ due to resonant interaction with the graphene plasmons. These decay rates are well below the frequencies at which the doping level can be modulated with currently available electronics ($\sim$ GHz), which is in turn much smaller than the plasmon frequency $\omega_p$ and the decay rate $\Gamma_{p}$. With this choice of parameters, we ensure that the emitter-plasmon interaction remains in the weak coupling regime, instantaneously following any modulation of the doping level. Under such conditions we can trace out the plasmonic degrees of freedom, and therefore the dynamics of the quantum emitter is completely described by the reduced density matrix $\rho$, whose temporal evolution is given by \cite{FT02}
\begin{eqnarray}
\frac{d\rho}{dt}=\frac{i}{\hbar}\left[\rho,\mathcal{H}\right]+ \frac{\Gamma_{11}}{2}\left[2\sigma\rho\sigma^{\dag}-\sigma^{\dag}\sigma\rho-\rho\sigma^{\dag}\sigma\right],
\label{1}\end{eqnarray}
where $\sigma$ is the annihilation operator of the emitter excited state, and the Hamiltonian reduces to $\mathcal{H}=\hbar\omega_0 \sigma^{\dag} \sigma $.

Fig.\ \ref{Fig1}(c) shows the normalized decay rate $\Gamma_{11}/\Gamma_0$ as a function of emission energy $\hbar\omega_0$ for two different doping levels corresponding to Fermi energies of $0.3$ and $0.4$\,eV, respectively. Then, choosing an emitter of transition energy $\approx0.22$\,eV (highest peak of black curve), we can switch the normalized decay rate from $\Gamma^{\rm on}_{11}/\Gamma_0>10^4$ down to $\Gamma^{\rm off}_{11}/\Gamma_0\sim10$, just by shifting the graphene doping level from $0.3$ to $0.4$\,eV. This clearly shows the feasibility of controlling the temporal evolution of the quantum emitter by modulating the doping level of the graphene nanodisk between on- and off-resonance conditions. We explore this possibility in more detail in Figs.\ \ref{Fig1}(d-e) for three different doping modulation profiles: rectangular (red curve), triangular (blue curve) and gaussian (green curve). The emitter is initially prepared in the excited state, and we study the plasmon generation rate $R=\left\langle\Gamma_{11}\sigma^{\dag}\sigma\right\rangle$, normalized to $\Gamma_{11}^{\rm on}$. This magnitude measures the number of plasmons generated per unit time, and it is calculated here neglecting decay channels others than plasmon generation. This assumption is well justified by the large values of $\Gamma^{\rm on/off}_{11}/\Gamma_0$. As shown in Fig.\ \ref{Fig1}(e), each different doping profile results in a totally different evolution of the quantum emitter, which reflects the complete temporal control achievable with the system under study. Actually, it is not difficult to obtain the analytical relation existing between $R$ and the single-emitter decay rate. Assuming the emitter in its excited state at time $t=t_0$, this relation reduces to
\begin{eqnarray}
\Gamma_{11}\left(t\right) = \frac{R\left(t\right)}{1-\int_{t_0}^{t}dt' R\left(t'\right)}.
\nonumber\end{eqnarray}
With the only constraint that $\Gamma_{11}\left(t\right)\le\Gamma_{11}^{\rm on}$, a desired profile $R\left(t\right)$ can be achieved with the temporal evolution of $\Gamma_{11}\left(t\right)$ prescribed by this equation, which is in turn obtained by directly modulating the doping voltage, and therefore $E_{\rm F}$, over time using the Lorentzian dependence of $\Gamma_{11}$ on $E_{\rm F}$ discussed in detail in the Supplemental Material (SM) \cite{superEPAPS}.

\begin{figure*}
\includegraphics[width=160mm,angle=0]{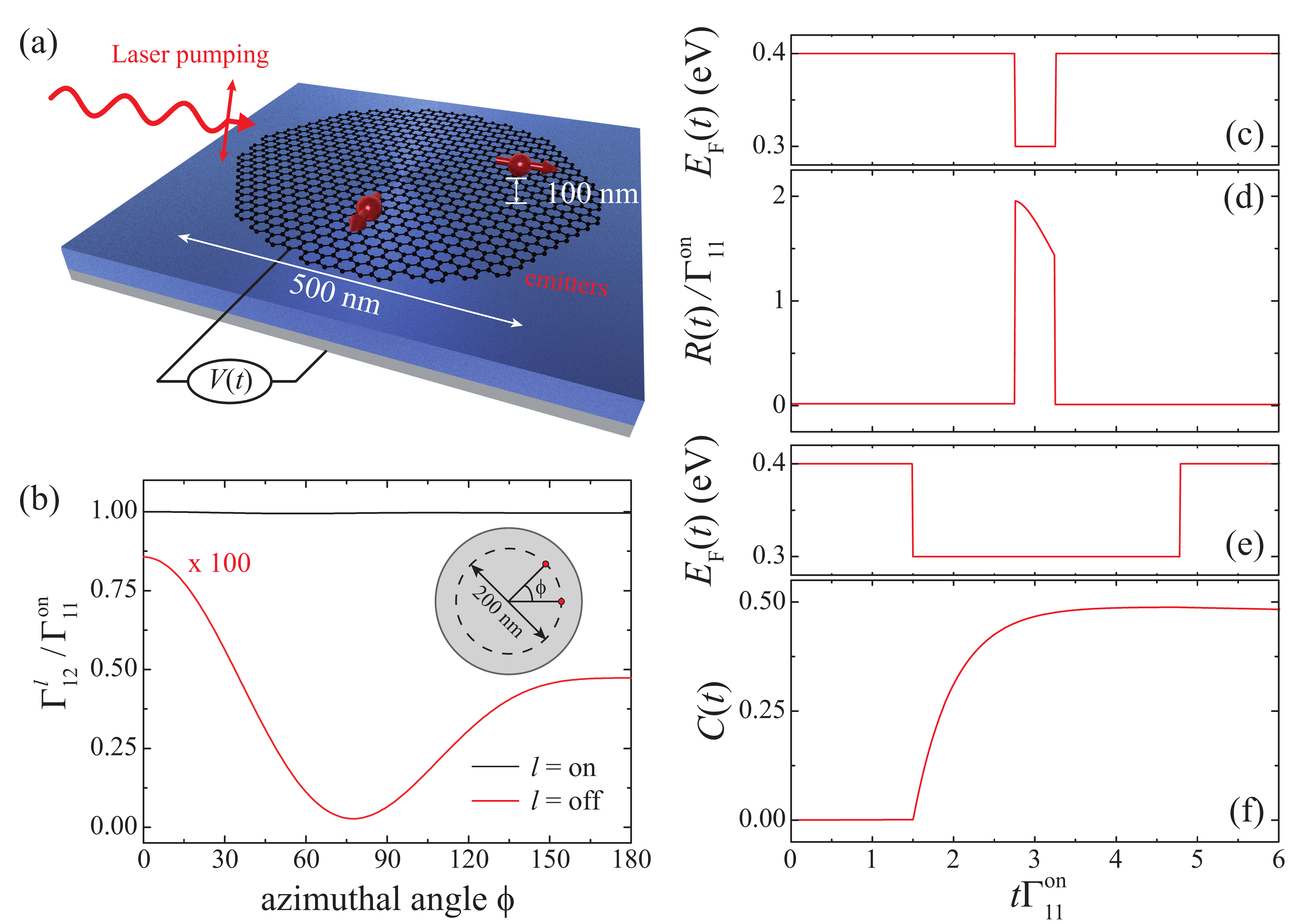}
\caption{Temporal control over the interaction between quantum dots mediated by graphene. (a) Two emitters are excited and their decay and mutual interaction is modulated electrically through the plasmons of a neighboring graphene disk. (b) Interaction rate $\Gamma_{12}$ as a function of the azimuthal angle between the positions of the dots (see inset) when plasmons of $0.108$\,eV energy are switched on ($E_{\rm F}=0.3\,$eV) and off ($E_{\rm F}=0.4\,$eV), and the emitter dipoles are along radial directions. The interaction rate is normalized to the on-resonance single-emitter decay rate $\Gamma_{11}^{\rm on}$. The imposed temporal evolution of the doping (c) is used to control the emission rate (d) with the two emitters initially prepared in their excited states. When only one of the emitters is initially excited, the degree of entanglement for the doping profile of (e) is quantified through the Wootters concurrence (f).}
\label{Fig2}
\end{figure*}

When a second quantum emitter is placed close to the graphene, the interaction between the two emitters can be also controlled over time. We investigate this possibility by studying the system depicted in Fig.\ \ref{Fig2}(a), where two identical emitters are placed $100$\,nm above a graphene nanodisk of $500$\,nm in diameter. The emitters are separated by a distance of $200$\,nm and oriented along orthogonal radial directions, so that they can be independently excited by light plane waves linearly polarized along orthogonal directions. Their temporal evolution is determined by the generalization of Eq.\ (\ref{1}),
\begin{eqnarray}
\frac{d\rho}{dt}=\frac{i}{\hbar}\left[\rho,\mathcal{H}\right]+ \sum_{i,j=1}^2\frac{\Gamma_{ij}}{2}\left[2\sigma_i\rho\sigma_j^{\dag}-\sigma_i^{\dag}\sigma_j\rho-\rho\sigma_i^{\dag}\sigma_j\right],
\label{222}\end{eqnarray}
where $\mathcal{H}=\hbar\omega_0 \sum_{i=1}^2\sigma_i^{\dag} \sigma_i $ and $\Gamma_{12}=\Gamma_{21}$ is the interaction rate. This magnitude is plotted in Fig.\ \ref{Fig2}(b) as a function of the azimuthal angle between the emitters, normalized to the on-resonance single-emitter decay rate $\Gamma_{11}^{\rm on}$.
When the doping level of the graphene nanodisk matches the on-resonance value, $\Gamma_{12}$ remains nearly equal to $\Gamma^{\rm on}_{11}$ for all angles. In contrast, the normalized interaction rate $\Gamma_{12}/\Gamma^{\rm on}_{11}$ drops below $0.01$ when the doping is tuned to off-resonance conditions. Therefore, it is possible to switch on and off the interaction between the emitters. Fig.\ \ref{Fig2}(d) quantifies this possibility through the temporal evolution of the plasmon generation rate $R=\left\langle\sum_{i,j=1}^{2}\Gamma_{ij}\sigma_i^{\dag}\sigma_j\right\rangle$ associated with the doping profile shown in Fig.\ \ref{Fig2}(c). The two emitters are assumed to be initially prepared in the excited state. Without interaction, they decay rather slowly and independently, resulting in a negligible value of $R$. This situation changes dramatically when the doping level is switched to the on-resonance condition, so that the emitters interact strongly and decay faster, producing a sudden jump in the plasmon generation rate.

The high degree of control displayed by this system can be exploited to temporally modulate different properties of the quantum emitters, such as their degree of entanglement. In Fig.\ \ref{Fig2}(f) we plot the temporal evolution of the Wootters concurrence \cite{W98} associated with the doping profile of Fig.\ \ref{Fig2}(e) when only one of the emitters is initially excited. The concurrence directly measures the degree of entanglement (1 for a maximally entangled state). Our system is capable of reaching a value of $C$ close to 0.5, which is understandable because a single excited emitter can be re-written as a mixture of a symmetric (superradiant) state and anti-symmetric (subradiant) state, but the subradiant state lives for very long time, such that there is 50\% chance of creating a long-lived entangled state. Larger degree of entanglement can be achieved by various means, such as utilizing quantum emitters with more than two levels, heralded schemes based on photon detection, or coherent dipole-dipole interactions between the emitters.

A similar scheme allows the on-demand modulation of the emission in systems consisting of many emitters, in which superradiance \cite{D1954,GH1982} can be produced and controlled electrostatically \cite{superEPAPS}.

In summary, here we demonstrate that electrical modulation of the plasmon frequency in graphene provides an ingenious solution to achieve temporal control over the evolution of quantum emitters placed in the vicinity of a graphene nanostructure. This leads to a new paradigm in quantum information processing technologies and serves as a platform on which to test quantum phenomena controlled by means of externally applied, classical electrostatic potentials.

This work has been supported by the Spanish MICINN (MAT2010-14885 and Consolider NanoLight.es) and the European Commission (FP7-ICT-2009-4-248909-LIMA and FP7-ICT-2009-4-248855-N4E). DEC acknowledges support from Fundacio Privada Cellex Barcelona. A.M. acknowledges financial support through FPU from the Spanish ME.


\end{document}